\documentclass[aps,prd,twocolumn,showpacs,floatfix,nofootinbib,superscriptaddress]{revtex4-1}

\usepackage{amsmath}
\usepackage{amssymb}
\usepackage{graphicx}
\usepackage{hyperref}
\usepackage{color}
\usepackage{mathtools}
\usepackage{times}
\usepackage{bm}
\usepackage{comment}

\usepackage[normalem]{ulem}

\newcommand{\la}{\langle}
\newcommand{\ra}{\rangle}

\newcommand{\up}[1]{{\rm #1}}
\newcommand{\beeq}{\begin{equation}}
\newcommand{\eneq}{\end{equation}}
\newcommand{\AVE}[1]{\left\langle#1\right\rangle}
\newcommand{\para}{\parallel}
\newcommand{\mpc}{{\rm Mpc}}
\newcommand{\hmpc}{{h^{-1}\mpc}}
\newcommand{\kvec}{{\bf k}}
\newcommand{\Nang}{\hat{\bm{n}}}
\newcommand{\ga}{\gamma}
\newcommand{\de}{\delta}
\newcommand{\HH}{\mathcal{H}}   
\newcommand{\rbar}{\bar r}       

\newcommand{\ngobs}{n_g^\up{obs}}
\newcommand{\ngobsb}{\bar n_{g,\Omega}^\up{obs}}

\newcommand{\den}{\de n}
\newcommand{\degi}{\de_g}
\newcommand{\deobs}{\de_g^\up{obs}}

\newcommand{\REVISED}[1]{#1}

\begin{document}

\title{Monopole Fluctuations in Galaxy Surveys}

\author{Jaiyul Yoo}
\email{jaiyul.yoo@uzh.ch}
\affiliation{Center for Theoretical Astrophysics and Cosmology,
Department of Astrophysics,
University of Z\"urich, Winterthurerstrasse 190,
CH-8057, Z\"urich, Switzerland}
\affiliation{Department of Physics, University of Z\"urich,
Winterthurerstrasse 190, CH-8057, Z\"urich, Switzerland}

\author{Daniel Eisenstein}
\email{deisenstein@cfa.harvard.edu}
\affiliation{Harvard-Smithsonian Center for Astrophysics,
60 Garden Street, Cambridge, MA 02138, USA}

\date{\today}

\begin{abstract}
 Galaxy clustering provides a powerful way to probe cosmology. This requires
understanding of the background mean density of galaxy samples, which is
 estimated from the survey itself by averaging the observed galaxy number 
density over the angular position. The angle average includes not only
the background mean density, but also the monopole fluctuation at each
redshift. Here for the first time we compute the monopole fluctuations 
in galaxy surveys and investigate their impact on galaxy clustering.
The monopole fluctuations vary as a function of
redshift, and it is correlated with other fluctuations, affecting
the two-point correlation function measurements. In an idealized
all-sky survey, the rms fluctuation
at $z=0.5$ can be as large as 7\% of the two-point correlation function
in amplitude at the BAO scale, and it becomes smaller than 1\% at $z>2$.
The monopole fluctuations are unavoidable, but they can be modeled.
We discuss its relation to the integral constraint and 
the implications for the galaxy clustering analysis.
\end{abstract}


\maketitle

\section{Introduction}
The expansion history of the Universe is one of the key ingredients
for understanding the energy contents of the Universe, and one of the best ways
to achieve this goal is to measure the distance-redshift relation of 
cosmological probes such as the cosmic microwave background (CMB)
temperature anisotropies
and galaxy clustering. The baryonic acoustic oscillation
(BAO) signal, which arises from the coupling of baryons and photons
in the early Universe \cite{PEYU70,SUZE70a}, is a standard ruler that 
can be used to measure the angular diameter distances to the 
last scattering surface and to the galaxy surveys. 
In particular, with the discovery of the late-time cosmic acceleration
\cite{RIFIET98,PEADET99}, precise measurements of the expansion history
in the late Universe are one of the primary goals in the recent and
upcoming large-scale galaxy surveys \cite{YOADET00,CODAET01,EIWEET11,
DESI13,LSST04,EUCLID11,WFIRST12},  and the BAO signals in galaxy
clustering have been detected with ever increasing precision
\cite{EIZEET05,BLCOET07,PEREET10,ANAUET12,ATBAET18,BAPAET21,DESIbao24}
(see \cite{EIHU98,EISEWH07,WEMOET13} for recent reviews).

For analyzing galaxy clustering, the background mean number density
of the galaxy samples should be subtracted, before the BAO signals
can be measured. Without full understanding of complex process of
galaxy formation and evolution, the mean number density is estimated
from the survey itself, which is then modulated by fluctuations of 
wavelength larger than the survey volume. The survey volume at each redshift 
is limited by a full-sky coverage, and the fluctuation over the entire
sky is called the {\it monopole} fluctuation, indistinguishable 
from the background mean value. In this {\it Letter}, for the first time 
we compute the monopole fluctuations in galaxy surveys and study
the impact of the monopole fluctuations on the two-point correlation
function at the BAO scales.

\section{Observed Mean and the Integral Constraint}
Here we briefly describe the standard procedure to estimate the observed
mean number density and analyze the galaxy number density fluctuation,
which is subject to the integral constraint \cite{PEEBL80,PENI91}.
The observed galaxy number density at the observed redshift~$z$ and
angle~$\Nang$ can be written as
\beeq
\ngobs(z,\Nang)=\bar n_g(z)\Big[1+\degi(z,\Nang)\Big]~,
\eneq
where $\bar n_g(z)$ is the number density in a background universe
and $\degi(z,\Nang)$ is the fluctuation of the observed galaxy number density.
The galaxy number density fluctuation~$\degi$ at the observed redshift
is mainly driven by the matter density fluctuation \cite{KAISE84,BBKS86} 
and the redshift-space distortion \cite{KAISE87}, 
in addition to the gravitational lensing effect \cite{NARAY89}
and other relativistic effects \cite{SAWO67}. The full relativistic
expression for~$\degi$ has been derived and shown to be gauge-invariant 
\cite{YOFIZA09,YOO10,BODU11}.
The goal is to compare various statistics of the galaxy
number density fluctuation~$\de_g$ to the measurements in galaxy surveys.
However, we do not know a priori the background number 
density $\bar n_g(z)$, and hence we cannot directly measure~$\de_g$.
 This is in contrast to the cases \cite{YOMIET19T,YOMIET19,BAYO21}
for the CMB temperature anisotropies
or the matter density fluctuations, in which we know their background
redshift evolution $\bar T\propto(1+z)$ or $\bar\rho_m\propto(1+z)^3$
and their values today ($\Omega_\ga$ and $\Omega_m$) are
part of a cosmological model with the adopted values for all the
cosmological parameters (including $\Omega_\ga$
and $\Omega_m$).\footnote{This could also be possible if we were to 
predict~$\bar n_g(z)$ based on, for instance, the Press-Schechter
formalism \cite{PRSC74,BCEK91,BBKS86}. 
However, given the uncertainty in the model,
we do not pursue this possibility here.}

Without a priori knowledge on the background number density and its 
redshift evolution, the observers use the survey data to measure the
observed mean~$\ngobsb(z)$ by simply averaging the observed galaxy number
density~$\ngobs(z,\Nang)$ over the observed angle~$\Nang$ as a function
of redshift~$z$ (see, e.g., \cite{EIANET01,COEIET08,WHBLET11,DESIbao24}):
\beeq
\ngobsb(z):=\int{d^2\hat n\over4\pi}
~\ngobs(z,\Nang)=\bar n_g(z)\Big[1+\den(z)\Big]~,
\eneq
where we defined the dimensionless angle-averaged galaxy fluctuation~$\den(z)$
(or the {\it monopole} fluctuation) as 
\beeq
\label{mono}
\den(z):=\int{d^2\hat n\over4\pi}~\degi(z,\Nang)~.
\eneq
Mind that only $\ngobsb(z)$ is an observable, not $\bar n_g(z)$ or $\den(z)$
separately. 
For simplicity, we have assumed an {\it idealized full-sky} survey and ignored
technical difficulties in practice such as a non-uniform angular selection 
function throughout the paper.

Hence the observed galaxy number density can be re-arranged  in terms of
the observed mean as
\beeq
\ngobs(z,\Nang)=\ngobsb(z)~{1+\degi(z,\Nang)\over1+\den(z)} ~,
\eneq
and the observed galaxy fluctuation is then
\beeq
\deobs(z,\Nang):={\ngobs(z,\Nang)\over\ngobsb(z)}-1=
{\degi(z,\Nang)-\den(z)\over1+\den(z)} ~,
\eneq
different from the standard theoretical prediction~$\degi(z,\Nang)$.
Expanding to the linear order in perturbation, we obtain the expression
\beeq
\label{dobs}
\deobs(z,\Nang)\approx\degi(z,\Nang)-\den(z)~,
\eneq
for the observed galaxy fluctuation we will use in this work.
Naturally, the observed galaxy fluctuation is subject to the integral
constraint at a given redshift:
\beeq
\label{INT}
0=\int~d^2\hat n~\deobs(z,\Nang)~~~~~\up{for}~\forall~z\in\up{survey}~,
\eneq
while the standard galaxy fluctuation~$\degi(z,\Nang)$ is {\it not}
subject to the same integral constraint:
\beeq
0\neq\int~d^2\hat n~\degi(z,\Nang)~.
\eneq
Note that the standard integral constraint in \cite{PENI91}
is formulated in terms of average 
over the survey volume, rather than average over the angle at each redshift.
Our equation~\eqref{INT} would correspond to the radial integral constraint
in \cite{DERU19}. 

\section{Monopole Fluctuations}
Our task now is to compute the monopole fluctuations in Equation~\eqref{mono}
as a function of redshift. In the past, little attention was paid to the
monopole fluctuations, as gauge issues in the monopole fluctuations 
result in infrared divergences and the monopole fluctuation cannot be separated
from the background mean value \cite{ZISC08}. First, 
the presence of infrared divergences in
cosmological observables such as the luminosity distance, CMB anisotropies
implies the deficiency in the theoretical descriptions, and it was shown
\cite{BIYO16,BIYO17,SCYOBI18,GRSCET20,BAYO21}
 that fully relativistic gauge-invariant theoretical descriptions
of the cosmological observables resolve the issues. Regarding the latter, 
can we measure the monopole fluctuations? Yes, they can be measured separately 
\cite{YOMIET19T,BAYO21},
if the background evolution is known (e.g., the matter density, the CMB 
temperature). Though this is not the case for galaxy clustering, its impact
is present in the galaxy $N$-point statistics. Here for the first time
we compute the monopole fluctuations in galaxy surveys.

To investigate the monopole fluctuations, we 
decompose the expression for the galaxy fluctuation~$\degi$ 
 in terms of spherical harmonics~$Y_{lm}(\Nang)$ as
\beeq
\label{decomp}
\degi(z,\Nang)=\sum_{lm}a_{lm}(z)Y_{lm}(\Nang)~,
\eneq
and the angular power spectrum is
\beeq
\label{power}
C_l(z_1,z_2)=\AVE{a_{lm}(z_1)a_{lm}^*(z_2)}~.
\eneq
The observed galaxy  fluctuation~$\deobs$ can be decomposed in the same way,
and its angular multipoles~$a_{lm}^\up{obs}$ are related to the angular 
multipoles~$a_{lm}$ of~$\de_g$ in Equation~\eqref{decomp} as
\beeq
a_{lm}^\up{obs}(z)={a_{lm}(z)\over1+\den(z)} \qquad \up{for}~~l\geq 1~.
\eneq
Since $\ngobsb(z)$ that includes the monopole fluctuation 
is defined as the observed mean,
the observed galaxy fluctuation~$\de_g^\up{obs}$ has {\it no}
monopole fluctuation:
\beeq
a_{00}^\up{obs}=0~,
\eneq
exactly in the same way the observed CMB temperature from 
the COBE FIRAS measurements \cite{FIRAS94,FICHET96,FIXSE09} 
includes the background temperature
and the monopole fluctuation.
The monopole fluctuation in Equation~\eqref{mono} is related to~$a_{00}$ as
\beeq
\den(z)={1\over \sqrt{4\pi}}~a_{00}(z)~.
\eneq

Using Equations~\eqref{decomp} and~\eqref{power}, the monopole power can 
be written as
\beeq
\label{powmon}
C_0(z_1,z_2)=4\pi\int d\ln k~\Delta^2_{\cal R}(k)~{\cal T}_0(k,z_1)
{\cal T}_0(k,z_2)~,
\eneq
where $\Delta^2_{\cal R}(k)=A_s(k/k_0)^{n_s-1}$ is the dimensionless
scale-invariant power spectrum of the comoving-gauge curvature 
perturbation~$\cal R$ at the initial time and ${\cal T}_0(k,z)$
is the monopole
transfer function for the galaxy fluctuation~$\degi$ at redshift~$z$.
For numerical computation, we assume the standard $\Lambda$CDM model,
in which the primordial fluctuation amplitude $A_s=2.1\times10^{-9}$, 
the spectral index $n_s=0.966$, the Hubble parameter $h=0.6732$, 
consistent with the
best-fit parameters from the Planck collaboration 
\cite{PLANCKcos18,PLANCKover18}.
To a good approximation, we can compute the monopole transfer function
by accounting for the matter density fluctuation~$\de_m$ with the galaxy bias 
factor~$b$ and the redshift-space distortion from the line-of-sight
velocity $V_\para$:
\beeq
\label{monoTF}
{\cal T}_0(k,z)=b~{\cal T}_m(k,z)j_0(k\rbar_z)
+{k^2\over\HH_z}{\cal T}_{v_m}(k,z)j_0''(k\rbar_z)~,
\eneq
where $j_0(x)$ is the spherical Bessel function, $\rbar_z$ is the comoving
distance to the redshift~$z$, $\HH_z$ is the conformal Hubble parameter,
${\cal T}_m$ and ${\cal T}_{v_m}$ are the transfer functions
for the (comoving-gauge) matter density fluctuation and the 
(Newtonian-gauge) velocity potential ($V_\para=:-\Nang\cdot\nabla v_m$).

\begin{figure}[t]
\centering
\includegraphics[width=0.48\textwidth]{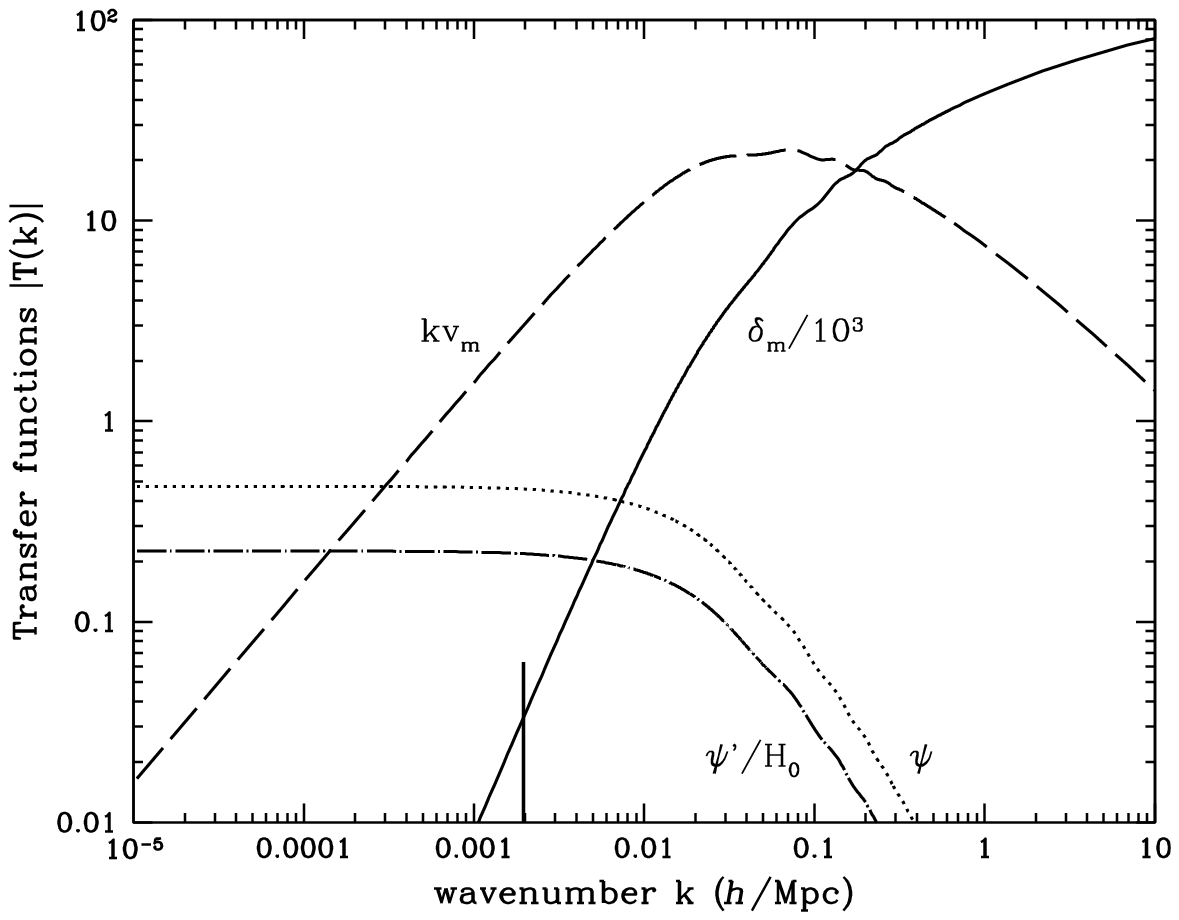}
\caption{Transfer functions ${\cal T}_i(k)$ for the individual 
perturbations~$\de p_i$ defined as $
\de p_i(\kvec,z)={\cal T}_i(k,z){\cal R}(\kvec)$.
Various curves show the transfer functions at redshift~$z=0$
for the matter density fluctuation~$\de_m$ (solid), the velocity 
potential~$v_m$
(dashed), the gravitational potential~$\psi$ (dotted), and the time-derivative
of the gravitational potential (dot dashed). Mind that ${\cal T}_m$ is
scaled by $10^3$ to fit in the plot. The short vertical line
indicates the wavenumber that corresponds to the first 
peak in $j_0(k\rbar_z)$ at redshift~$z=1$.}
\label{TF}
\end{figure}

\begin{figure}[t]
\centering
\includegraphics[width=0.48\textwidth]{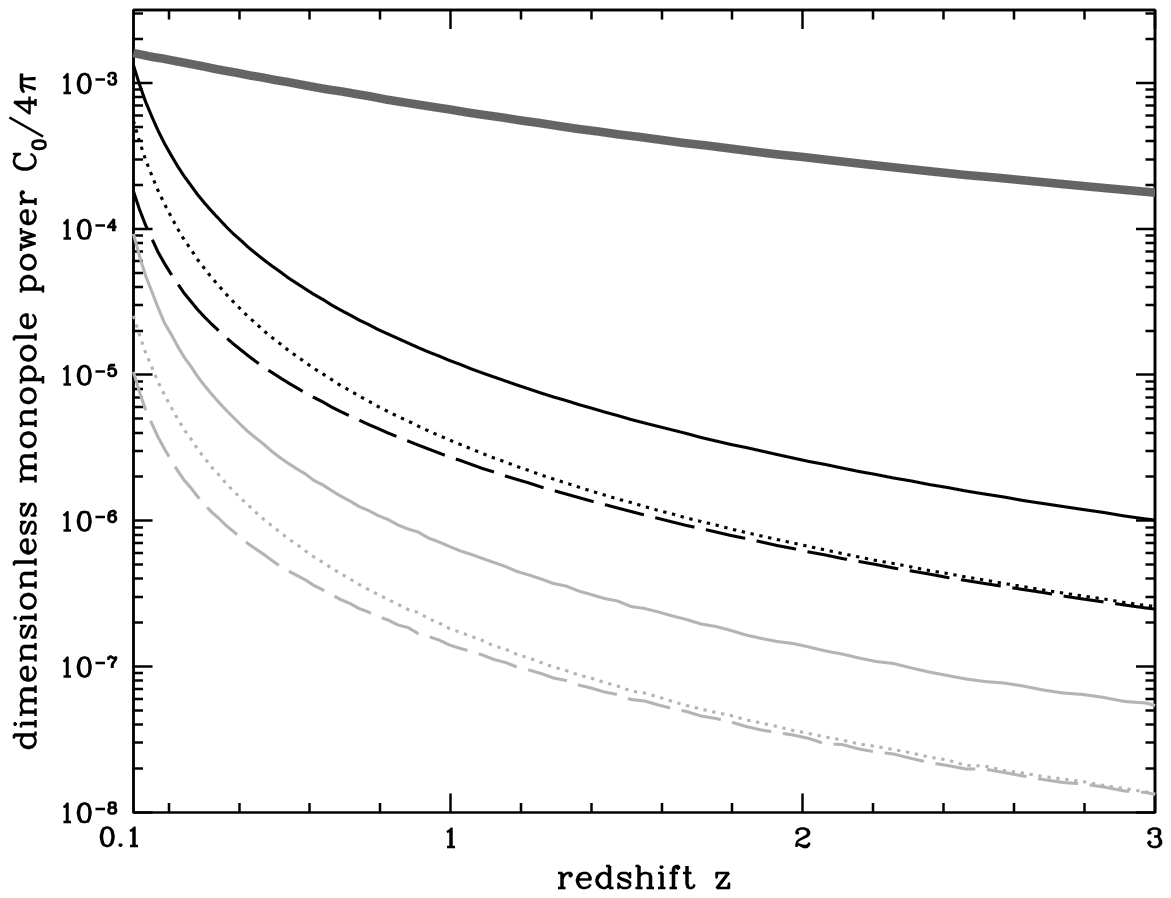}
\caption{Monopole power $C_0(z)/4\pi$ and the impact 
 on the two-point correlation function. Thick solid curve shows 
the amplitude of the two-point matter density
correlation function at the BAO peak position.
Black curves (solid, dotted, dashed) 
show the monopole power $C_0(z)/4\pi$ as a function of 
redshift~$z$ from the matter density fluctuation (dotted), 
the redshift-space distortion (dashed), and their sum (solid). 
Gray curves (solid, dotted, dashed)
show $C_0(z_1,z_2)/4\pi$ as a function of redshift~$z$,
with two redshifts~$z_1$, $z_2$ of the galaxy pair, 
corresponding to the line-of-sight separation $r=100~\hmpc$,
 and with redshift~$z$ corresponding to the half the separation.}
\label{Fig:pow}
\end{figure}

Figure~\ref{TF} shows the transfer functions of the individual perturbations
in the galaxy fluctuation~$\de_g$ at redshift~$z$. The transfer functions
are defined as $\de p_i(\kvec,z)=:{\cal T}_i(k,z){\cal R}(\kvec)$
in terms of the initial fluctuation~${\cal R}$.
The matter density fluctuation (solid in Figure~\ref{TF}, the first
term in Eq.~[\ref{monoTF}]) is the dominant contribution to
galaxy clustering \cite{KAISE84},
and the other contributions such as the line-of-sight
velocity (dashed) and the gravitational potential (dotted) are smaller
by orders of magnitude. The vertical line indicates the scale at $z=1$, 
beyond which
the contributions of the individual transfer functions are further 
suppressed due to the spherical
Bessel function~$j_0(x)$ in the monopole transfer function~${\cal T}_0(k,z)$. 
Note that the transfer function slope for the matter density fluctuation
asymptotically reaches $\sim0.34$ on small scales. 
On large scales ($x\ll1$), the transfer functions for the 
gravitational potential are constant, responsible for infrared divergences
in the monopole fluctuations. Their contributions, however, collectively cancel 
on large scales and remain small, if a correct relativistic
formula is used \cite{JESCHI12,SCYOBI18,GRSCET20,MIYOMA23,MAYO23}.
Note that there is no gravitational lensing
contribution~$\kappa$ in the monopole transfer function.
The second term in Equation~\eqref{monoTF} is the redshift-space distortion
\cite{KAISE87}, 
or the spatial derivative of the line-of-sight velocity, which can
be re-arranged as $-f~{\cal T}_m(k,z)j_0''(k\rbar_z)$ by using the
Einstein equation with the logarithmic growth rate~$f$.

The (black) dotted curve in Figure~\ref{Fig:pow}
shows the monopole power~$C_0$ scaled with~$4\pi$
from the matter density fluctuation~$\de_m$ (or a constant
bias factor $b=1$ for the galaxy sample).
Given that the initial condition~$\Delta^2_{\cal R}(k)$ is nearly 
scale-invariant ($n_s\approx1$),  
the monopole power in Equation~\eqref{powmon} can be
read off from the transfer functions in Figure~\ref{TF} at the peak
of the spherical Bessel function. The matter density fluctuation decreases
with increasing redshift, as the decreasing
growth factor~$D(z)$ reduces the overall
amplitude of the transfer function. Furthermore, with larger comoving 
distance~$\rbar_z$, the contributions of the individual transfer functions
are shifted to a larger characteristic scale $k\sim1/\rbar_z$, 
further reducing the monopole power at higher redshift.
The black dashed curve represents the contribution of the redshift-space 
distortion. Though modulated differently
with~$j_0''(x)$, it closely follows the shape
of the matter density power, especially at high redshift, where
$f\simeq1$. Finally, the solid curve shows the full monopole power
from the matter density fluctuation and the redshift-space distortion
in Equation~\ref{monoTF}. Note that both contributions oscillate with different
periods and the monopole power represents the combined
transfer functions at the peak.
For our base model ($b=1$), the monopole power $C_0/4\pi\approx10^{-3}$
at $z=0.1$, but decreases rapidly below $10^{-5}$ beyond $z=1$.

\section{Impact on the Two-Point Correlation Function}
Having computed the monopole power as a function of redshift, we are now
in a position to quantify the impact of the monopole fluctuations 
on the two-point correlation function measurements. Given the configuration
of two galaxies and the observed galaxy fluctuation in Equation~\eqref{dobs},
we compute the ensemble average, or the two-point correlation function:
\beeq
\label{CORR}
\left\la\deobs(\bm{x}_1)\deobs(\bm{x}_2)\right\ra=
\left\la \degi(\bm{x}_1)\degi(\bm{x}_2)\right\ra
-\left\la \den(z_1)\den(z_2)\right\ra~,~~~~~~
\eneq
where we shortened the notation $\bm{x}=(z,\Nang)$. We can show that
three additional terms in the observed two-point correlation function
are indeed identical to the angular monopole power~$C_0(z_1,z_2)$:
\beeq
\left\la \den(z_1)\den(z_2)\right\ra
=\left\la \den(z_1)\degi(\bm{x}_2)\right\ra={1\over4\pi}C_0(z_1,z_2)~.
\eneq
Equation~\eqref{CORR} can be better expressed as
\beeq
\label{CORR2}
\xi_g^\up{obs}(r)=\xi_g(r)-{1\over4\pi}~C_0(z_1,z_2)~,
\eneq
such that the correlation function~$\xi_g^\up{obs}(r)$ we measure in surveys 
is the correlation function~$\xi_g(r)$
we predict with~$\degi$ with the monopole
power~$C_0(z_1,z_2)/4\pi$ taken out, where $r$~is the separation between
two galaxy positions. 

Equation~\eqref{CORR} is valid, as both hand sides are ensemble averaged.
The ensemble averages in practice can be replaced by averaging over many
galaxy pairs in the same configurations in surveys.
However, the monopole fluctuations
that appear in Equation~\eqref{dobs} or~\eqref{CORR} before ensemble
average correspond to a {\it single} realization of random fluctuations at each
redshift. Nevertheless, we used the ensemble average to compute its
contribution to~$\xi_g^\up{obs}(r)$.
With these caveats, here
we use Equation~\eqref{CORR2} to investigate
the impact of the monopole fluctuations on the measurements of the observed
two-point correlation function.
As shown in Figure~\ref{Fig:pow},
the monopole power is $C_0(z)/4\pi\simeq10^{-3}$ at $z=0.1$ and decreasing
fast at higher redshift (solid), if galaxies are simply modeled
as the matter fluctuation. Hence the impact is negligible, whenever
$\xi_g\gg10^{-3}$, which is the case for
 most of the dynamic range of the correlation
function measurements. However, the monopole fluctuations may be relevant
on large scales such as the BAO scale, where $\xi_g$ is small.

While the monopole power is just a function
of two redshifts~$z_1$ and~$z_2$
of the observed galaxies, independent of their angular
positions, it affects {\it not only} the observed correlation function along the
line-of-sight direction, {\it but also} the observed correlation function 
along the transverse direction. 
First, for all the configurations of two galaxies
at the same redshift~$z$ (or along the transverse direction),
the monopole power remains unchanged~$C_0(z)/4\pi$,
regardless of their physical separation~$r$ set by two angular 
directions~$\Nang_1$ and~$\Nang_2$ at the same redshift~$z$. Hence, 
the correction is a constant
shift in the transverse correlation function, but this shift is a function
of redshift, as shown in Figure~\ref{Fig:pow} (black curves).
Second, for a configuration of two galaxies 
along the line-of-sight direction
(same~$\Nang$, but two different redshifts~$z_1$ and~$z_2$), the monopole
power $C_0(z_1,z_2)/4\pi$ in this case is directly a function of their 
separation~$r$, as the difference in the comoving distances~$\rbar_z$ at
two redshifts is directly related to the separation~$r$.

Gray curves in Figure~\ref{Fig:pow} show the monopole power
$C_0(z_1,z_2)/4\pi$ along the line-of-sight direction for two galaxies
located at~$z_1$ and~$z_2$, with the corresponding separation~$r=100~\hmpc$. 
The dotted and dashed curves represent the matter density fluctuation and
the redshift-space distortion, while the solid curve is the combination.
With destructive interference of two spherical Bessel functions from two
different distances~$\rbar_z$, the monopole power at two different redshifts 
with $r=100\hmpc$
is negative and smaller in the absolute amplitude than the monopole power 
at the same redshift.

The thick solid curve in Figure~\ref{Fig:pow} shows the amplitude of 
the matter density two-point correlation function $\xi_m(r)$
at the BAO peak position. As described, the exact correction
 $C_0(z_1,z_2)/4\pi$ to be made to~$\xi_g(r)$ depends on the configuration 
of two galaxy pairs, but here we make a simple estimate, leaving the
detailed investigation for a future work \cite{MAYOEI24}.
At redshift $z=0.5$, the monopole power affects the amplitude of 
two-point correlation function at the BAO peak position by 7\% along the
transverse direction or smaller along the line-of-sight direction.
This ratio decreases at higher redshift to 1\% at $z=2$ along the
transverse direction. In our simple model for galaxies ($b=1$),
while the ratio is independent of the growth factor~$D(z)$, it
decreases at high redshift due to the increase in characteristic scale 
$k\sim1/\rbar_z$ of~$C_0(z)$.  However, we stress that our results
are obtained in a full-sky survey.

\section{Conclusion and Discussion}
The galaxy mean number density needs to be subtracted
for galaxy clustering analysis, 
and without {\it ab initio} knowledge of galaxy evolution,
the mean number density is estimated from the survey itself. Since the observed
mean contains the monopole fluctuation at each redshift, the observed
galaxy two-point correlation function is affected by the monopole fluctuations.
For the same reason, any $N$-point statistics such as the three-point
correlation function will be affected by the monopole fluctuations. 
Since the monopole fluctuation~$\den\sim\sqrt{C_0}$ is small 
($C_0\approx10^{-3}$ at $z=0.1$ and $C_0\approx10^{-5}$ at $z=1$),
the impact of the monopole fluctuation is, however, limited to large scales 
such as the BAO scale, where the two-point correlation function is small.
Assuming the rms value for the monopole fluctuations, we find that 
the corrections to the two-point correlation function can be as large
as 7\% at $z=0.5$ and 1\% at $z=2$ in the amplitude at the BAO scale
for a survey with full sky coverage. 
\REVISED{This level
of correction at the BAO scale is larger than the other systematic errors
such as the nonlinear evolution and gravitational lensing
\cite{SMSCSH08,SEET10,YOMI10,SHZA12}.
Furthermore, the level of corrections stays relatively high even at~$z\geq1$,
as the monopole corrections arise directly from the galaxy clustering itself.}

For simplicity, we have assumed a constant bias factor~$b=1$ for the
galaxy sample and ignored nonlinearity in the transfer functions.
Different values of the galaxy bias factor and its time evolution will
certainly change the ratio of the corrections from the monopole fluctuations
to the two-point correlation function, albeit not significantly.
Despite the suppression from the spherical Bessel function,
nonlinearity in the transfer functions can enhance the monopole fluctuations
by boosting the transfer function on small scales, 
in particular at low redshift, where the nonlinearity is significant
 and the characteristic scale $k\sim1/\rbar_z$ approaches the nonlinear scale.
\REVISED{More importantly, the nonlinearity in galaxy clustering would
boost the galaxy correlation function~$\xi_g(r)$ and the
monopole correction~$C_0(z_1,z_2)/4\pi$ in almost the same way
that the change in the ratio of the monopole correction to the
galaxy correlation at the BAO scale remains small.}

The monopole fluctuations represent the fundamental limitation to
our theoretical modeling of the two-point correlation function,
or the {\it cosmic variance}. They cannot be removed even 
in idealized surveys with infinite volume, 
as we only have access to a  single light cone
volume and there exists only one realization of the monopole fluctuations
at each redshift. Without full Euclidean average including translation
of the observer position \cite{MIYOET20},
the ensemble average cannot be replaced by 
spatial average, and any measurements of random fluctuations
are inevitably limited by the cosmic variance (see, e.g., 
\cite{PEEBL80,PEACO99,DODEL03}).
 While the galaxy evolution
should be locally smooth and close to a passive evolution $(\bar n_g\propto
1/a^3)$ if limited
to a small redshift bin~$\Delta z$, it would remain difficult to separate,
as the monopole fluctuations~$\den(z)$ are small, i.e., 1\% in~$\bar n_g(z)$
at $z=1$. 

In practice, the observed mean is often estimated 
by either the spline fit to the redshift distribution or
shuffling the redshift measurements in surveys (see \cite{SAPERA12}
for details), which can mitigate some bias arising 
from galaxy clustering. This clustering would correspond to the corrections
from higher angular multipole fluctuations at each redshift, but the monopole
fluctuations cannot be removed by shuffling the angular positions.
For the same reason, when the sky coverage is incomplete, subsequent angular 
multipoles such as the dipole fluctuations and so on can act
as the monopole fluctuations in the full sky, as those low angular multipole
fluctuations at each redshift are again indistinguishable from the mean number
density with a partial sky coverage. The ``monopole'' fluctuation
defined in Equation~\eqref{mono} will depend {\it not only} on~$C_0$,
{\it but also} all~$C_l$ with $l\geq0$ in surveys with incomplete sky coverage.
\REVISED{Consequently, the contributions from these higher multipoles
would greatly enhance the ``monopole'' fluctuation in case of incomplete
sky coverage, but its impact on galaxy clustering will require further
numerical studies beyond the scope of current work.}
Complicated angular selection functions
in real surveys such as holes, disjoint patches would also affect the
observed mean number density.

The observed correlation function in galaxy surveys
is analyzed by further averaging over the angle of the separation vector
for a galaxy pair, such as the monopole correlation~$\xi_0(r)$,
the quadrupole~$\xi_2(r)$, and the hexadecapole~$\xi_4(r)$ 
\cite{HAMIL92,COFIWE94} (see \cite{HAMAET03,RESAET12,BESEET17}
for recent measurements). While the
monopole fluctuation~$\den(z)$ is independent of angular separation,
it depends on redshift, such that the effects of the monopole
fluctuations on measurements of the multipole correlation functions~$\xi_l(r)$ 
are non-trivial and requires further investigations \cite{MAYOEI24}.
\REVISED{Odd multipoles such as
the dipole~$\xi_1(r)$ can exist in galaxy clustering
\cite{MCDON09,BOHUGA14},
if two different galaxy populations are used,
as exchanging two galaxies at~$z_1$ and~$z_2$
results in a different configuration. The monopole power~$C_0(z_1,z_2)/4\pi$
in the observed correlation function would also contribute to the odd
multipoles, but we suspect that its contribution remains small and it is
difficult to isolate the monopole corrections from the odd multipoles.}
In contrast, the power spectrum analysis is non-local by nature, 
and the integral constraint is always part of the power spectrum 
analysis \cite{PENI91,FEKAPE94,VOSZ96,TETAHE97}.
Hence we suspect that the impact on the power spectrum
analysis is likely to be small, though there might be tangible impact
again on the redshift-space multipole power spectra \cite{DERU19} due to 
the integral constraint specified in Equation~\eqref{INT},
rather than integration over the volume in the standard power spectrum
analysis.

\REVISED{The BAO peak position measured in galaxy surveys is a standard ruler,
by which we infer the angular diameter distances to the galaxy
samples and constrain cosmological parameters 
\cite{EIZEET05,BLCOET07,PEREET10,ANAUET12,ATBAET18,BAPAET21,DESIbao24}.
Its measurements yield two parameters~$\alpha_\para$
and~$\alpha_\perp$ that characterize the BAO peak position at each redshift,
 compared to the position predicted in the adopted fiducial cosmology.
A percent level shift in those parameters due to the monopole corrections
would translate into a percent level systematic error in the Hubble parameter
beyond the level of precision the upcoming surveys like the DESI aim for.}
Measurements of the BAO
peak position in the correlation function are, however, performed by
first marginalizing over the smooth power around the peak position
due to the nonlinearity and scale-dependent galaxy bias 
\cite{SESIET08,DESIsys24}.
Hence, while the monopole fluctuations are expected to affect the measurements
of the BAO peak position especially at low redshift, its precise impact 
after the marginalization process
requires further investigations beyond the scope of this work.
\REVISED{Despite this caveat, we emphasize that the level of corrections due to
the monopole fluctuations is at least an {\it order-of-magnitude}
larger over the redshift range $z=0.5-2.5$ than any other effects
around the BAO scales such as the gravitational lensing and the nonlinear
evolution of matter and bias \cite{SMSCSH08,SEET10,SHZA12}.
}
\\

\section*{Acknowledgments}
We acknowledge useful discussions with Yan-Chuan Cai.
This work is supported by the Swiss National Science Foundation
and a Consolidator Grant of the European Research Council.

\bibliography{monopole.bbl}

\end{document}